\documentclass[%
reprint,
onecolumn,
 amsmath,amssymb,
 aps,
]{revtex4-2}

\usepackage{graphicx}
\usepackage{dcolumn}
\usepackage{bm}
\usepackage{hyperref}
\usepackage[mathlines]{lineno}
\usepackage{mathtools}
\usepackage{xcolor}

\begin{document}

\title{Relational GNS Fusion and an Idempotent Gauge-Gravity Fixed Point}

\author{Gerard J. Verbiest}
\affiliation{Faculty of Mechanical, Maritime and Materials Engineering; Department of Precision and Microsystems Engineering, Delft University of Technology, 2628 CD, Delft, The Netherlands}
\email{G.J.Verbiest@tudelft.nl}

\begin{abstract}
We propose a relational noncommutative framework in which quantum kinematics, semiclassical geometry, internal gauge structure, and ultraviolet fixed-point behaviour arise as sectors of a common $*$-algebra equipped with a positive state and its Gelfand--Naimark--Segal (GNS) representation. The central construction is a projected fusion product on normalized quadratic current operators. Within the closed current sector, $P_{\rm rel}(O_A^2)=2O_A$ up to irrelevant corrections, and ultraviolet self-similarity expressed as $X_A\star X_A=X_A$ with $X_A=r_AO_A$ yields the interacting normalized fixed point $r_A^*=1/2$. With the retained operator basis fixed by the GNS Gram metric, a reduced per-generation fermionic current convention for the gauge sectors and canonical graviton normalization give a common algebraic value for $(10/3)g_Y^2$, $2g_2^2$, $2g_3^2$, and $32\pi G\mu^2$. We discuss conditional emergence of Lorentzian geometry and Einstein gravity, a Standard-Model-like internal algebra, three generation channels from a minimal quartic pairing sector, minimal selection of a $3+1$ dimensional branch, and relational time with a branch-dependent arrow of records. A Standard Model--Pati--Salam renormalization-group analysis provides a phenomenological consistency test of the proposed ultraviolet basin.
\end{abstract}

\maketitle

% ============================================================
\section{Introduction}
\label{sec:introduction}
% ============================================================

Quantum field theory and general relativity are formulated in conceptually different languages: the former assumes quantum states and fields on spacetime, while the latter makes spacetime geometry dynamical. Algebraic quantum theory provides a useful starting point because a positive state on a $C^*$-algebra defines a Hilbert-space representation through the GNS construction \cite{Landsman1998,FewsterRejzner2020}. Noncommutative-geometric approaches further show that finite internal algebras can encode Standard-Model gauge and matter structure and admit Pati--Salam extensions \cite{ChamseddineConnes2007,ChamseddineConnes2010,ChamseddineConnesvanSuijlekom2013}.

We investigate whether quantum kinematics, geometry, and internal interactions can instead be viewed as effective sectors of a more primitive relational noncommutative algebra $\mathcal A$ equipped with a positive normalized state $\omega$, with $\omega(A^*A)\ge0$ and $\omega(\mathbf1)=1$. The GNS representation $(\mathcal H_\omega,\pi_\omega,|\Omega\rangle)$ satisfies $\omega(A)=\langle\Omega|\pi_\omega(A)|\Omega\rangle$.

The second ingredient is a state-dependent projected product
\begin{equation}
    A\star B:=P_{\rm rel}(AB),
    \label{eq:intro_star}
\end{equation}
where $P_{\rm rel}$ retains the collective operator sector relevant at the scale under consideration. For normalized quadratic current operators $O_A$, the closed current sector will be shown to satisfy
\begin{equation}
    O_A\star O_A=2O_A+\mathcal O_{\rm irr}.
    \label{eq:intro_fusion}
\end{equation}
Requiring the physical interaction $X_A=r_AO_A$ to reproduce itself under fusion,
\begin{equation}
    X_A\star X_A=X_A,
    \label{eq:intro_idempotency}
\end{equation}
gives the interacting algebraic fixed point $r_A^*=1/2$.

The numerical fixed-point coordinate depends on how the retained operators are normalized, so that convention must be fixed before different sectors can be compared. Throughout this work the GNS Gram metric fixes the normalization of the retained operators. For gauge currents we use a reduced per-generation fermionic current metric, with a common replica multiplicity divided out before defining the collective normalized current. In this convention one Standard Model generation gives $K_Y=10/3$ and $K_2=K_3=2$. The Pati--Salam fermion multiplets $(2,1,4)+(1,2,\bar4)$ similarly give $K_L=K_R=K_4=2$ per generation; scalar fields affect the RG flow but are not included in this reduced fermionic current metric. On the gravitational side the canonical Einstein--Hilbert vertex convention fixes $\kappa^2=32\pi G$. We therefore define
\begin{equation}
    r_Y=\frac{10}{3}g_Y^2,\qquad r_2=2g_2^2,\qquad r_3=2g_3^2,\qquad r_G=32\pi G\mu^2,
    \label{eq:intro_strengths}
\end{equation}
and, in the same fixed normalization convention, the ultraviolet relation
\begin{equation}
    \frac{10}{3}g_Y^{*2}=2g_2^{*2}=2g_3^{*2}=32\pi G^*\mu^2=\frac12.
    \label{eq:intro_master}
\end{equation}
This is not conventional equality of gauge couplings. It is equality of normalized interaction strengths in the specified GNS-normalized operator basis; an arbitrary rescaling of the retained operators would change the numerical coordinate of the idempotent and is therefore not an allowed additional convention once the Gram-metric normalization has been fixed.

\begin{figure}[t]
    \centering
    \includegraphics[width=0.98\textwidth]{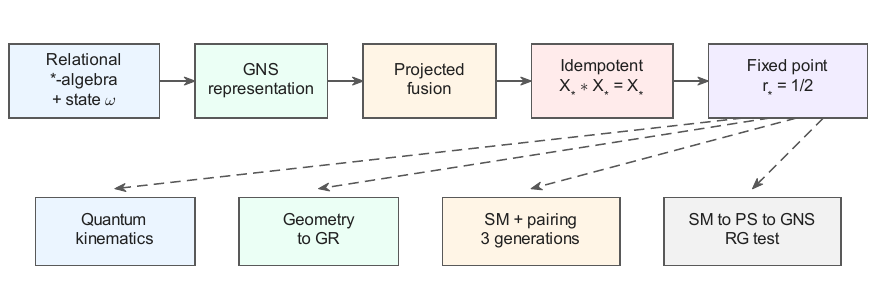}
    \caption{Logical architecture of the relational GNS/fusion framework. The diagram is a dependency map; not all arrows have equal logical status.}
    \label{fig:framework}
\end{figure}

The logical status of subsequent claims is deliberately separated. The fusion idempotent is derived within a specified closed current sector. Emergent geometry, $D=4$, three generations, and relational time require additional semiclassical or minimality assumptions. Standard Model--Pati--Salam running is used only as a consistency test. Figure~\ref{fig:framework} summarizes this architecture.

% ============================================================
\section{Relational $*$-algebra and GNS construction}
\label{sec:gns_construction}
% ============================================================

\subsection{Algebra, state, and GNS representation}
\label{subsec:gns_algebra}

The microscopic kinematics is a unital noncommutative $*$-algebra $\mathcal A$ with associative product and involution $(AB)^*=B^*A^*$. Relational loop generators may be represented by unitaries $U_\gamma$ satisfying $U_\gamma^*=U_{\gamma^{-1}}$. Noncommutativity can be modeled locally by a Weyl relation $UV=e^{i\theta}VU$, without assuming a fundamental spacetime manifold \cite{RieffelSchwarz1998,BordemannWaldmann1996}.

A positive normalized state $\omega$ defines the null ideal $\mathcal N_\omega=\{A:\omega(A^*A)=0\}$. The quotient $\mathcal A/\mathcal N_\omega$ carries the inner product
\begin{equation}
    \langle[A],[B]\rangle_\omega=\omega(A^*B),
    \label{eq:gns_inner}
\end{equation}
whose completion is $\mathcal H_\omega$. Left multiplication defines $\pi_\omega(A)[B]=[AB]$, while the class of the unit gives the cyclic vector $|\Omega_\omega\rangle$. These are standard consequences of the GNS theorem \cite{Landsman1998,FewsterRejzner2020}.

The physical hypothesis added here is that state-dependent collective sectors of this representation can acquire quantum, geometric, and internal interpretations. For centered relational operators $\delta A_I=A_I-\omega(A_I)\mathbf1$, the correlation matrix $C_{IJ}=\omega(\delta A_I^*\delta A_J)$ identifies independent collective directions. A semiclassical sector is one in which selected collective observables have small relative fluctuations and sufficiently clustered connected correlations.

\subsection{Quantum kinematics from noncommutative relations}
\label{subsec:gns_quantum_kinematics}

Writing a local Weyl pair as $U=e^{i\epsilon Q}$ and $V=e^{i\epsilon P}$, with $\theta=\epsilon^2\hbar_{\rm eff}+\mathcal O(\epsilon^3)$, the Baker--Campbell--Hausdorff expansion gives
\begin{equation}
    [Q,P]=i\hbar_{\rm eff}\mathbf1+\mathcal O(\epsilon),
    \label{eq:gns_ccr}
\end{equation}
and therefore $\Delta_\omega Q\,\Delta_\omega P\ge\hbar_{\rm eff}/2+\mathcal O(\epsilon)$. The framework assumes a single central deformation scale, schematically $\Theta_{IJ}=\hbar_{\rm eff}\Omega_{IJ}$, rather than attempting to predict the numerical value of a dimensionful Planck constant.

The GNS construction supplies kinematics but no fundamental external time. Relational time is introduced later by conditioning on a semiclassical clock sector.

% ============================================================
\section{Projected fusion algebra}
\label{sec:fusion_algebra}
% ============================================================

\subsection{Projection and fusion tensor}
\label{subsec:fusion_projection}

Let $\mathcal V_{\rm rel}=\mathrm{span}\{O_I\}$ be the retained centered operator subspace with GNS Gram matrix $G_{IJ}=\omega(O_I^*O_J)$. Orthogonal projection gives
\begin{equation}
    P_{\rm rel}(A)=O_I(G^{-1})^{IJ}(O_J,A)_\omega,
    \label{eq:fusion_projector}
\end{equation}
with summation over repeated basis indices. Defining $A\star B=P_{\rm rel}(AB)$, the structure constants are
\begin{equation}
    O_J\star O_K=C_{JK}{}^I O_I,\qquad C_{JK}{}^I=(G^{-1})^{IL}\omega(O_L^*O_JO_K).
    \label{eq:fusion_tensor}
\end{equation}
Projected products of composite operators and operator mixing are familiar in OPE and renormalization-group constructions, although the present fusion interpretation is specific to this framework \cite{Hollands2023,PaganiSonoda2018,PolonyiSailer2001}.

\subsection{Normalized quadratic currents}
\label{subsec:fusion_currents}

For a current sector $A$, normalize $J_A^a$ so that $\omega(\widehat J_A^a\widehat J_A^b)=\delta^{ab}$. When several identical fermion generations are present, the common replica factor is first removed from the reduced current metric; this fixes the collective basis used for cross-sector comparisons and prevents a common multiplicity from being counted both in the operator normalization and in $r_A$. We then define the centered quadratic singlet
\begin{equation}
    O_A=\frac12\left[\widehat J_A^a\widehat J_A^a-\omega(\widehat J_A^a\widehat J_A^a)\mathbf1\right].
    \label{eq:fusion_O}
\end{equation}
If this is the only relevant quadratic singlet, $P_2(O_A^2)=C_AO_A$, where
\begin{equation}
    C_A=\frac{\omega(O_A^3)}{\omega(O_A^2)}.
    \label{eq:fusion_C}
\end{equation}
For $N$ orthonormal quasi-free current components, $\omega(O^2)=N/2$ and $\omega(O^3)=N$, hence $C=2$ independently of $N$.

The quasi-free condition is sufficient but not necessary. For one normalized $J\to-J$ symmetric variable with fourth and sixth cumulants $\kappa_4$ and $\kappa_6$,
\begin{equation}
    C=2+\delta C,\qquad \delta C=\frac{8\kappa_4+\kappa_6}{4+2\kappa_4}.
    \label{eq:fusion_cumulants}
\end{equation}
Under block averaging, cluster-summable connected cumulants scale as $\kappa_n\sim b^{1-n/2}$ and vanish for $n>2$. Conserved currents in a scale-invariant $D$-dimensional sector have $\Delta_J=D-1$, which is compatible with infrared summability in $D=4$ \cite{Gillioz2022,Stanev2013}.

For a simple current algebra the invariant quadratic singlet is unique up to normalization. The self-product decomposes schematically as $O^2=A\mathbf1+BO+Q_4$, where centering removes the identity and $Q_4$ denotes higher current composites. If these are irrelevant under $P_{\rm rel}$, the closed current sector satisfies
\begin{equation}
    \boxed{O_A\star O_A=2O_A+\mathcal O_{\rm irr}.}
    \label{eq:fusion_main}
\end{equation}
Orthogonal current sectors have vanishing quadratic cross-projection in the minimal truncation.

The leading possible obstruction is an external singlet $S_s$ with the same full quantum numbers,
\begin{equation}
    O_A\star O_A=(2+\delta C_A)O_A+\sum_s\epsilon_{As}S_s+\mathcal O_{\rm irr}.
    \label{eq:fusion_corrected}
\end{equation}
Symmetry can force individual mixing coefficients to vanish; otherwise they are calculable from the full GNS fusion tensor.

% ============================================================
\section{Idempotent ultraviolet fixed point}
\label{sec:uv_fixedpoint}
% ============================================================

\subsection{Algebraic fixed point and normalization}
\label{subsec:uv_idempotent}

For $X_A=r_AO_A$, idempotency $X_A\star X_A=X_A$ together with Eq.~(\ref{eq:fusion_main}) gives $2r_A^2=r_A$, hence
\begin{equation}
    r_A^*=0\quad\text{or}\quad \boxed{r_A^*=\frac12}.
    \label{eq:uv_fixedpoint}
\end{equation}
Equivalently, $\Pi_A=O_A/2$ is an idempotent of the closed projected algebra. This differs conceptually from defining the UV fixed point solely by $\beta_i(g^*)=0$, although non-Gaussian fixed points are central to asymptotic-safety and gauge--Yukawa programmes \cite{Reuter1998,DaumHarstReuter2010,LitimSannino2014,LitimMojazaSannino2016}.

For the reduced Standard Model fermionic current traces and the canonical graviton vertex normalization,
\begin{equation}
    \boxed{\frac{10}{3}g_Y^{*2}=2g_2^{*2}=2g_3^{*2}=32\pi G^*\mu^2=\frac12,}
    \label{eq:uv_master}
\end{equation}
corresponding in this convention to $g_Y^*=\sqrt{3/20}$, $g_2^*=g_3^*=1/2$, and $G^*\mu^2=1/(64\pi)$. The value $1/2$ is therefore a coordinate of the idempotent in a fixed operator basis, not an invariant under arbitrary rescalings $O_A\to c_AO_A$. The GNS Gram metric fixes that basis on the algebraic side, while the canonical Einstein--Hilbert normalization fixes the gravitational vertex. In the Pati--Salam fermionic sector used below, the same reduced convention gives $K_L=K_R=K_4=2$, so the normalized quantities above $M_R$ are $r_L=2g_L^2$, $r_R=2g_R^2$, and $r_4=2g_4^2$.

If only the self-fusion coefficient is shifted, the interacting solution becomes $r_A^*=1/(2+\delta C_A)$. External singlet mixing instead requires solving the idempotency condition in the enlarged operator space.

\subsection{Weak universality and crossover}
\label{subsec:uv_unification}

Sector-by-sector $r_A^*=1/2$ is a weak universality statement. A stronger master-unification claim requires anisotropic perturbations between sectors to be more irrelevant than the common direction. For two sectors, $O_-=(O_1-O_2)/\sqrt2$ and its susceptibility $\chi_-=\int d^Dx\,\langle O_-(x)O_-(0)\rangle_c$ provide a concrete diagnostic: finite susceptibility is consistent with suppression of anisotropic block averages, but does not by itself constitute a full proof of a unique master direction.

The fixed point is asymptotic rather than a finite-scale boundary condition. A minimal illustrative interpolation is $dr/d\ln\mu=\omega_{\rm RG}r(1-2r)$. For gravity, choosing $\omega_{\rm RG}=2$ and matching to $G_N$ gives
\begin{equation}
    r_G(\mu)=\frac12\frac{\mu^2}{\mu^2+\Lambda_{\rm GNS}^2},\qquad \Lambda_{\rm GNS}^2=\frac{1}{64\pi G_N}.
    \label{eq:uv_crossover}
\end{equation}
Thus $r_G(\Lambda_{\rm GNS})=1/4$ and $r_G\to1/2$ only asymptotically. The interpolation is illustrative; the stability exponent must ultimately be derived from the coarse-graining operator.

% ============================================================
\section{Emergent geometry and infrared gravity}
\label{sec:emergent_gravity}
% ============================================================

\subsection{Correlation geometry and massless spin--2}
\label{subsec:gravity_geometry}

Geometry is associated with a semiclassical GNS branch rather than with the abstract algebra alone. For relational observables $J_I^a$, assume a low-rank correlation factorization
\begin{equation}
    C_{IJ}^{ab}=e_I{}^ae_J{}^b\eta_{ab}+\Delta C_{IJ}^{ab}.
    \label{eq:gravity_factorization}
\end{equation}
When relational directions admit a continuum identification $I\to\mu$, the collective frame defines
\begin{equation}
    g_{\mu\nu}=e_\mu{}^ae_\nu{}^b\eta_{ab}.
    \label{eq:gravity_metric}
\end{equation}
The frame is locally non-unique under $e_\mu{}^a\to\Lambda^a{}_b(x)e_\mu{}^b$, giving an emergent local Lorentz redundancy. A Lorentzian signature is a property of the selected branch, not of the positive GNS Hilbert norm.

Relational loop holonomies admit the usual small-loop expansion $U_\gamma=\mathbf1+F_{\mu\nu}\Sigma^{\mu\nu}+\mathcal O(\Sigma^2)$, providing effective connection and curvature data in the continuum sector.

For $g_{\mu\nu}=\bar g_{\mu\nu}+\kappa h_{\mu\nu}$, preservation of relational gauge redundancy is assumed to induce the linearized transformation
\begin{equation}
    h_{\mu\nu}\to h_{\mu\nu}+\bar\nabla_\mu\xi_\nu+\bar\nabla_\nu\xi_\mu.
    \label{eq:gravity_diffeo}
\end{equation}
This protects a massless spin--2 mode; failure of the coarse-grained constraint algebra could instead generate additional or massive degrees of freedom \cite{Carlip2014}.

\subsection{Universal coupling and Einstein infrared dynamics}
\label{subsec:gravity_ir}

Assuming that the emergent massless spin--2 sector couples consistently to the conserved total stress tensor of the matter sectors represented in the same state-dependent geometry, the leading linear interaction has the universal form
\begin{equation}
    \mathcal L_{\rm int}=\frac{\kappa}{2}h_{\mu\nu}T^{\mu\nu}.
    \label{eq:gravity_universal}
\end{equation}
This universality is therefore a consistency requirement on the emergent gravitational sector rather than a consequence of common geometric representation alone. Consistency of a Lorentz-invariant massless spin--2 field likewise requires universal coupling to conserved energy--momentum \cite{Weinberg1964,Deser1970}.

At energies below the microscopic scale the most general local diffeomorphism-invariant action is an EFT derivative expansion,
\begin{equation}
    S_{\rm eff}=\int d^4x\sqrt{-g}\left[-\Lambda_{\rm eff}+\frac{M_{\rm eff}^2}{2}R+c_1R^2+c_2R_{\mu\nu}R^{\mu\nu}+\cdots\right].
    \label{eq:gravity_eft}
\end{equation}
Therefore Einstein--Hilbert gravity is the leading infrared dynamics, with higher-derivative corrections suppressed by the ultraviolet scale \cite{Donoghue1994,Donoghue2022}. The claim is conditional: the framework must still demonstrate a stable Lorentzian frame and anomaly-free massless spin--2 constraint algebra microscopically.

% ============================================================
\section{Internal algebra and Standard Model structure}
\label{sec:standard_model}
% ============================================================

The low-energy internal sector is taken to contain the finite algebra
\begin{equation}
    \mathcal A_F=\mathbb C\oplus\mathbb H\oplus M_3(\mathbb C),
    \label{eq:sm_algebra}
\end{equation}
which is the familiar algebra of the noncommutative-geometric Standard Model \cite{ChamseddineConnes2007,ChamseddineConnes2008,ChamseddineConnes2010}. Its factors provide an Abelian phase, an $SU(2)$ doublet structure, and color $U(3)$, with unimodularity/anomaly constraints reducing the redundant Abelian degree of freedom and yielding the effective Standard Model gauge group \cite{AlvarezGraciaBondiaMartin1995}.

For one generation we use $Q_L,L_L,u_R,d_R,e_R$ and optionally $\nu_R$. With $Y_H=1/2$, gauge invariance of the Yukawa terms gives $Y_u=Y_Q+Y_H$, $Y_d=Y_Q-Y_H$, $Y_e=Y_L-Y_H$, and $Y_\nu=Y_L+Y_H$. Anomaly cancellation includes $3Y_Q+Y_L=0$ and the standard cubic and mixed gravitational conditions \cite{LohitsiriTong2019}. The Standard Model branch then gives
\begin{equation}
    Y_Q=\frac16,\quad Y_L=-\frac12,\quad Y_u=\frac23,\quad Y_d=-\frac13,\quad Y_e=-1,\quad Y_\nu=0.
    \label{eq:sm_hypercharges}
\end{equation}
The resulting reduced per-generation fermionic traces are $K_Y=10/3$ and $K_2=K_3=2$, giving the normalization used in Eq.~(\ref{eq:uv_master}). Repeating the representation over three generations multiplies the unreduced traces by the same common factor; in the reduced collective-current convention used here that replica factor is divided out before defining $r_A$.

The Pati--Salam enlargement
\begin{equation}
    SU(4)_C\times SU(2)_L\times SU(2)_R
    \label{eq:sm_ps}
\end{equation}
arises naturally in related noncommutative-geometric constructions \cite{ChamseddineConnesvanSuijlekom2013}. For one fermion family $(2,1,4)+(1,2,\bar4)$, the reduced quadratic indices are $K_L=K_R=K_4=2$: for example, $K_L=4T(\mathbf2)=2$, while the two $SU(4)$ fundamentals each contribute $2T(\mathbf4)=1$. With $Y=T_R^3+(B-L)/2$, the tree-level matching condition is $g_Y^{-2}=g_R^{-2}+(2/3)g_4^{-2}$. Here $\mathcal A_F$ is used as a minimal empirically adequate low-energy internal algebra, not as a uniqueness theorem excluding larger microscopic algebras.

% ============================================================
\section{Dimensional and generation selection}
\label{sec:dimensions_generations}
% ============================================================

\subsection{Minimal spacetime dimension}
\label{subsec:dimensions}

Define the effective dimension of a semiclassical branch by the rank of its retained correlation structure, $D_{\rm eff}=\mathrm{rank}\,C_{\rm rel}$. A massless spin--2 field in $D$ spacetime dimensions has
\begin{equation}
    N_{\rm grav}=\frac{D(D-3)}{2}
    \label{eq:dim_graviton}
\end{equation}
local degrees of freedom, so propagating Einstein gravity requires $D\ge4$ \cite{Tekin2003}. Chiral Weyl representations require even $D$, but chirality alone does not select four dimensions because higher even dimensions also admit Weyl spinors \cite{Park2022Clifford,Traubenberg2005}. Hence
\begin{equation}
    D_{\rm min}=4
    \label{eq:dim_four}
\end{equation}
is the lowest branch simultaneously supporting local massless spin--2 propagation and chirality. With one Lorentzian timelike direction this gives three spatial dimensions. Higher-dimensional branches are not excluded; they are non-minimal extensions.

\subsection{Minimal quartic pairing and three generations}
\label{subsec:generations}

The finite internal algebra fixes one-generation gauge representation content but does not explain the observed generation count \cite{ChamseddineConnes2008,Stephan2009}. Consider instead the minimal connected four-leg fusion sector. Four labels admit exactly three perfect matchings,
\begin{equation}
    P_1=(12)(34),\qquad P_2=(13)(24),\qquad P_3=(14)(23),
    \label{eq:gen_pairings}
\end{equation}
so $N_{\rm pair}=(4-1)!!=3$. We identify the corresponding space $\mathcal H_{\rm gen}=\mathrm{span}\{|P_1\rangle,|P_2\rangle,|P_3\rangle\}$ as a candidate generation multiplicity space.

Permutations act on this space as $\mathbf3=\mathbf1\oplus\mathbf2$, where the singlet is proportional to $P_1+P_2+P_3$ and the relative doublet satisfies $x+y+z=0$. Gauge generators act as
\begin{equation}
    T_A=T_A^{\rm SM}\otimes I_3,
    \label{eq:gen_gauge}
\end{equation}
so all three channels carry identical gauge charges, while Yukawa/fusion operators may act non-trivially on $\mathcal H_{\rm gen}$. Different diagonalizing bases then produce the usual structures $V_{\rm CKM}=U_u^\dagger U_d$ and $U_{\rm PMNS}=U_e^\dagger U_\nu$.

The exact combinatorial result is the three-dimensional matching space. Its identification with the three observed fermion generations remains a physical hypothesis requiring a dynamical flavor calculation.

\begin{figure}[t]
    \centering
    \includegraphics[width=0.88\textwidth]{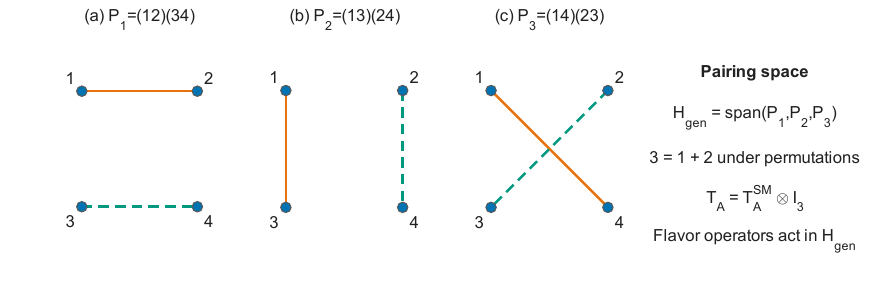}
    \caption{The three inequivalent perfect matchings of four relational labels. They span the proposed generation space, with gauge generators acting identically on all three channels.}
    \label{fig:generation_pairings}
\end{figure}

% ============================================================
\section{Relational time and the arrow of records}
\label{sec:relational_time}
% ============================================================

A fundamental state may satisfy a stationary constraint $\widehat H_{\rm tot}|\Psi\rangle=0$. In a semiclassical branch admitting an approximate clock--system split $\mathcal H_\omega\simeq\mathcal H_C\otimes\mathcal H_S$, conditional states are
\begin{equation}
    |\psi(\tau)\rangle_S\propto{}_C\langle\tau|\Psi\rangle.
    \label{eq:time_conditional}
\end{equation}
This is the Page--Wootters mechanism: non-trivial subsystem evolution can arise relative to an internal clock even when the global state is stationary \cite{PageWootters1983,MarlettoVedral2017}. If the clock constraint reduces semiclassically to $(P_\tau+H_S)|\Psi\rangle\simeq0$, one obtains
\begin{equation}
    i\hbar_{\rm eff}\partial_\tau|\psi(\tau)\rangle=H_S|\psi(\tau)\rangle.
    \label{eq:time_schrodinger}
\end{equation}
The clock split is not assumed to be arbitrary; admissible clocks must have suitable monotonicity, small fluctuations, weak back-reaction, and stable correlations. The uniqueness of such a clock sector remains open, and relational-time proposals are known to face clock-selection ambiguities \cite{Stoica2026ClockAmbiguity}.

The time arrow is treated separately. Decoherence produces robust pointer states and redundant environmental records \cite{Zurek2003,Zurek2009,PazZurek1993,RiedelZurekZwolak2014}. In a low-entropy semiclassical branch we define the arrow operationally as the orientation in which stable records accumulate,
\begin{equation}
    \mathcal R(\tau_2)\supseteq\mathcal R(\tau_1)\quad(\tau_2>\tau_1),\qquad \frac{dS_{\rm cg}}{d\tau}>0.
    \label{eq:time_arrow}
\end{equation}
The underlying algebra does not presently derive the low-entropy branch condition or a unique global arrow; the orientation is branch-relative \cite{Hartle2020}.

% ============================================================
\section{Gauge--gravity crossover and RG consistency}
\label{sec:rg_crossover}
% ============================================================

\subsection{Standard Model to Pati--Salam running}
\label{subsec:rg_running}

The algebraic fixed point is asymptotic, so the phenomenological test is whether measured couplings can enter a plausible neighbourhood of $\mathbf r_*=(1/2,1/2,1/2)$ before the perturbative spacetime description crosses over to the GNS regime.

We use the $\overline{\rm MS}$ reference point $Q_0=173.1\,\mathrm{GeV}$ with $g_Y=0.35853877$, $g_2=0.64765961$, $g_3=1.1636241$, and $y_t=0.93480082$ \cite{MartinRobertson2019}. Standard Model gauge running is known to high loop order \cite{MihailaSalomonSteinhauser2012}. For the benchmark breaking scale $M_R=10^{13}\,\mathrm{GeV}$, our integration gives the values shown in Table~\ref{tab:rg_results}. The choice of $M_R$ is one relevant symmetry-breaking parameter, not a prediction.

At $M_R$, $g_L=g_2$, $g_4=g_3$, and the hypercharge matching relation determines $g_R$. Above $M_R$ we use three generations of $(2,1,4)+(1,2,\bar4)$ and scalars $\phi=(2,2,1)$ and $\Delta=(1,2,4)$, in $(L,R,4)$ ordering. The one-loop coefficients are $b_L=-3$, $b_R=-7/3$, and $b_4=-31/3$. Evaluating the general two-loop gauge formulas \cite{LuoWangXiao2003,SartoreSchienbein2021} gives
\begin{equation}
    B=\begin{pmatrix}
    8&3&45/2\\
    3&50/3&75/2\\
    9/2&15/2&-206/3
    \end{pmatrix}.
    \label{eq:rg_B}
\end{equation}

\begin{table}[t]
\caption{Representative benchmark RG results. Values at $\Lambda_{\rm GNS}$ characterize the crossover neighbourhood rather than an exact finite-scale fixed point.}
\label{tab:rg_results}
\centering
\begin{tabular}{lccc}
\hline\hline
Quantity & one loop & two loop & interpretation\\
\hline
$g_Y(M_R)$ & 0.42128 & 0.42211 & SM\\
$g_2(M_R)$ & 0.54410 & 0.54556 & SM / $g_L$\\
$g_3(M_R)$ & 0.58367 & 0.58068 & SM / $g_4$\\
$g_R(M_R)$ & -- & 0.52448 & matched\\
$g_L(\Lambda_{\rm GNS})$ & 0.51356 & 0.51413 & PS\\
$g_R(\Lambda_{\rm GNS})$ & 0.50182 & 0.50271 & PS\\
$g_4(\Lambda_{\rm GNS})$ & 0.47390 & 0.47315 & PS\\
\hline\hline
\end{tabular}
\end{table}

A standalone Standard Model continuation to $\Lambda_{\rm GNS}\simeq8.6\times10^{17}\,\mathrm{GeV}$ gives approximately $r_2\simeq0.525$, $r_3\simeq0.503$, but $r_Y\simeq0.721$. Thus the non-Abelian sectors approach the normalized fixed-point neighbourhood whereas fundamental hypercharge does not, motivating its Pati--Salam embedding.

For the two-loop Pati--Salam result, the normalized values are $r_L\simeq0.529$, $r_R\simeq0.505$, and $r_4\simeq0.448$, with Euclidean distance $d_{\rm GNS}\simeq0.060$ from $(1/2,1/2,1/2)$. Gauge two-loop corrections remain at the per-mille level. Conservative Yukawa stress tests likewise shift the gauge couplings only at the few-per-mille level over the benchmark interval. A complete Pati--Salam Yukawa calculation remains to be done.

Heavy fields modify the matching through $g_Y^{-2}=g_R^{-2}+(2/3)g_4^{-2}+\Delta_{\rm th}$. Closing the benchmark inverse matching requires $\Delta_{\rm th}\simeq-0.0289$, about $-0.52\%$ of $1/g_Y^2$. The exact threshold must ultimately follow from the heavy spectrum rather than being treated as an arbitrary tuning parameter.

The intended sequence is therefore
\begin{equation}
    \text{perturbative SM}\to\text{Pati--Salam}\to\text{GNS crossover}\to X_*\star X_*=X_*,
    \label{eq:rg_chain}
\end{equation}
not exact finite-scale equality $r_A(\Lambda_{\rm GNS})=1/2$.

\begin{figure}[t]
    \centering
    \includegraphics[width=0.96\textwidth]{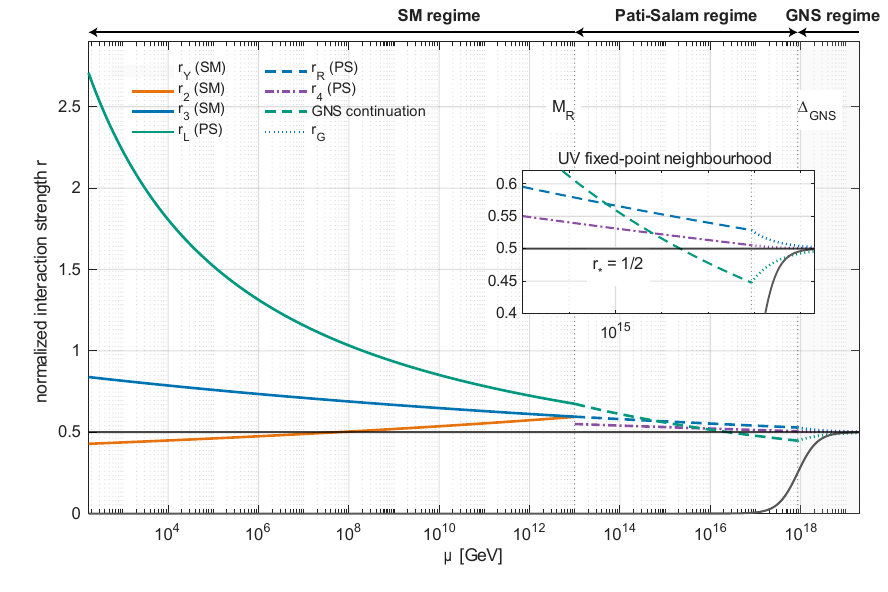}
    \caption{Gauge--gravity crossover on a logarithmic energy axis. Standard Model running is used below $M_R$, Pati--Salam running between $M_R$ and $\Lambda_{\rm GNS}$, and the high-energy shaded region denotes crossover into the non-perturbative GNS basin. The inset resolves the neighbourhood of the asymptotic line $r_*=1/2$; the dotted continuation beyond $\Lambda_{\rm GNS}$ is schematic rather than perturbative RG evolution.}
    \label{fig:rg_crossover}
\end{figure}

% ============================================================
\section{Falsifiability and limitations}
\label{sec:falsifiability}
% ============================================================

The framework contains algebraic results, conditional emergence statements, and phenomenological consistency checks. These should not be assigned the same logical status.

\subsection{Decisive tests}
\label{subsec:fals_tests}

The central fusion mechanism fails if unavoidable relevant singlet mixing destroys the interacting idempotent continuously connected to $r_A=1/2$. This is a microscopic test of the full fusion tensor in Eq.~(\ref{eq:fusion_tensor}).

The ultraviolet connection fails if a complete multi-loop Standard Model--Pati--Salam calculation, including Yukawa and threshold effects, cannot enter any reasonable neighbourhood of the proposed GNS basin. Separately, the symmetry-breaking construction loses economy if no single relevant deformation can reproduce consistent low-energy matching without a proliferation of independent scales or thresholds.

Strong master unification requires more than the common value $r_A^*=1/2$: anisotropic coupling modes must become irrelevant under coarse-graining. If they remain independently critical, weak fixed-point universality may survive while the stronger claim of a unique master interaction fails.

Conditional low-energy identifications have their own tests. The geometric sector must produce a stable rank-four Lorentzian frame and an anomaly-free massless spin--2 constraint algebra. The generation proposal must eventually reproduce realistic flavor dynamics intrinsically on $\mathcal H_{\rm gen}$ rather than introduce a separate unexplained flavor space. The relational-time sector still requires a physically selected clock and a low-entropy branch condition.

\subsection{Open calculations}
\label{subsec:fals_open}

The principal unresolved calculations are: a classification of external singlet mixing; the complete Pati--Salam heavy threshold spectrum and gauge--Yukawa running; microscopic determination of the breaking scale; flavor masses and mixing; an explicit rank-four Lorentzian semiclassical GNS state; and selection of the low-entropy cosmological branch. The present construction also does not attempt detailed models of dark matter, inflation, baryogenesis, or the cosmological constant.

\begin{table}[t]
\caption{Logical status of the main results.}
\label{tab:claim_status}
\centering
\begin{tabular}{lll}
\hline\hline
Result & status & principal condition\\
\hline
GNS representation & derived & positive normalized state\\
Projected fusion tensor & derived & chosen relevant space\\
$O_A\star O_A=2O_A$ & closed-sector result & clustering and singlet closure\\
$r_A^*=1/2$ & closed-sector result & idempotency\\
Common normalized value & weak universality & current normalization\\
Strong master direction & conjectural & anisotropic irrelevance\\
Lorentzian metric & conditional & frame factorization\\
Einstein IR dynamics & conditional/EFT & massless spin--2 and diffeomorphism symmetry\\
$D=4$ & minimality selection & spin--2 propagation and chirality\\
Three generations & structural identification & quartic pairing\\
Relational time & conditional & admissible clock\\
SM--PS--GNS trajectory & consistency test & one breaking scale and thresholds\\
\hline\hline
\end{tabular}
\end{table}

% ============================================================
\section{Discussion}
\label{sec:discussion}
% ============================================================

The distinguishing feature of the construction is the use of a projected GNS fusion idempotent as the ultraviolet organizing principle. Conventional asymptotic-safety and gauge--Yukawa approaches characterize ultraviolet completion by non-Gaussian zeros of beta functions \cite{Reuter1998,DaumHarstReuter2010,LitimSannino2014}; here the primary condition is instead $X_*\star X_*=X_*$, with RG evolution interpreted as an effective approach to that algebraic structure. The proposal is complementary rather than contradictory: the fusion algebra fixes a candidate normalized fixed-point value, while a microscopic coarse-graining calculation must still determine stability exponents and the complete flow field.

The internal matter sector borrows the successful finite algebra of noncommutative geometry but does not assume a fundamental almost-commutative spacetime. Likewise, the Pati--Salam regime has independent precedent in spectral and asymptotically safe constructions \cite{ChamseddineConnesvanSuijlekom2015,MolinaroSanninoWang2018}. The geometric sector is more conditional: once a Lorentzian frame, diffeomorphism redundancy, and a single universally coupled massless spin--2 mode emerge, Einstein gravity is the expected infrared EFT \cite{Carlip2014,Deser1970,Donoghue1994}. The microscopic challenge is to derive those prerequisites from the state-dependent GNS correlations.

Minimality rather than absolute uniqueness underlies the numerical selections. Four dimensions are the lowest branch supporting both local massless spin--2 propagation and chirality; higher even dimensions are not excluded. The three perfect matchings of the minimal quartic sector give a natural three-dimensional generation space, but its physical identification must be validated by flavor dynamics. Relational time similarly emerges conditionally from a suitable clock sector, while the arrow of records still depends on a low-entropy branch.

The quantitative RG test is deliberately modest. The benchmark trajectory shows no present obstruction to entering a GNS neighbourhood, and standalone hypercharge provides a non-trivial reason to introduce a non-Abelian parent. This is not an exact finite-scale unification claim and does not determine $M_R$.

The highest-priority calculations are therefore clear: enlarge the relevant GNS singlet basis and compute the full fusion tensor; determine the multi-sector stability spectrum; complete the Pati--Salam threshold and gauge--Yukawa analysis; derive a flavor operator on the pairing space; and construct an explicit semiclassical GNS state with a closed Lorentzian gravitational constraint algebra. These steps test the central algebra before extending the framework to additional phenomenology.

% ============================================================
\section{Conclusion}
\label{sec:conclusion}
% ============================================================

We have proposed a relational noncommutative framework in which a positive state and its GNS representation organize quantum, geometric, gauge, generation, and temporal sectors. The principal algebraic result is that, within a closed normalized-current sector, $O_A\star O_A=2O_A$ and fusion idempotency selects the interacting value $r_A^*=1/2$. With current-metric normalization this yields the common gauge--gravity relation in Eq.~(\ref{eq:uv_master}).

The remaining results are deliberately conditional. A suitable semiclassical correlation sector can support a Lorentzian metric and, if it contains an anomaly-free universally coupled massless spin--2 mode, Einstein gravity emerges as the leading infrared EFT. The finite algebra $\mathbb C\oplus\mathbb H\oplus M_3(\mathbb C)$ carries Standard-Model-like matter, while minimality gives $D_{\rm min}=4$ and a three-dimensional quartic pairing space as a candidate origin of generation multiplicity. A stationary global state can support relational Schr\"odinger evolution, with a branch-dependent arrow associated with stable records.

The benchmark Standard Model--Pati--Salam running is compatible, within current approximations, with entry into the proposed GNS basin. The theory remains testable: the idempotent must survive external singlet mixing, precision RG evolution must remain compatible with the basin without excessive tuning, anisotropic interaction modes must be suppressed for strong master unification, and the semiclassical state must reproduce the observed gravitational and flavor structures.

The proposal can be summarized compactly as
\begin{equation}
    \text{relational algebra}\to\text{GNS representation}\to\text{projected fusion}\to X_*\star X_*=X_*.
    \label{eq:conclusion_chain}
\end{equation}
Its viability ultimately depends on whether the remaining microscopic calculations preserve this algebraic structure while reproducing the observed low-energy world.

% ============================================================
% ACKNOWLEDGMENTS
% ============================================================

\section*{Acknowledgement}

The author acknowledges extensive iterative discussions with ChatGPT (OpenAI; GPT-5.6 Sol during the final revision stage), used as an interactive research and writing tool for conceptual exploration, critical questioning, mathematical cross-checking, literature organization, numerical-code development, and refinement of the presentation of the framework developed in this work. These discussions were particularly useful in separating derived results from conditional assumptions, identifying possible falsification criteria, testing alternative interpretations, and organizing the gauge--gravity, GNS fusion, emergent-geometry, generation, and relational-time sectors into a coherent manuscript.

ChatGPT was also used to assist with editing portions of the manuscript and preparing reproducibility scripts. All physical assumptions, mathematical claims, numerical results, literature selections, interpretations, and conclusions presented in this work were reviewed and remain the responsibility of the author.

% ============================================================
% DATA AVAILABILITY
% ============================================================

\section*{Data availability}

The numerical code used to reproduce the renormalization-group calculations and figures reported in this work is provided as supplementary material. No additional research data were generated or analysed.

% ============================================================
% REFERENCES
% ============================================================

\section*{References}

\bibliographystyle{iopart-num}
\bibliography{references}

\end{document}